\documentclass[twocolumn,prl,tightenlines,superscriptaddress]{revtex4-1}

\usepackage{amsmath}
\usepackage{amssymb,amsfonts,latexsym}
\usepackage{bm}
\usepackage[mathcal]{euscript}
\usepackage{graphicx}
\usepackage{epsfig}
\usepackage{color}
\usepackage{xfrac}
\usepackage{multirow}

\begin{document}

\title{Self-Propelled Rods: Linking Alignment-Dominated and Repulsion-Dominated Active Matter}

\author{Xia-qing Shi}
\affiliation{Center for Soft Condensed Matter Physics and Interdisciplinary Research, Soochow University, Suzhou 215006, China}
\affiliation{Service de Physique de l'Etat Condens\'e, CEA, CNRS, Universit\'e Paris-Saclay, CEA-Saclay, 91191 Gif-sur-Yvette, France}

\author{Hugues Chat\'{e}}
\affiliation{Service de Physique de l'Etat Condens\'e, CEA, CNRS, Universit\'e Paris-Saclay, CEA-Saclay, 91191 Gif-sur-Yvette, France}
\affiliation{Beijing Computational Science Research Center, Beijing 100094, China}
\affiliation{Center for Soft Condensed Matter Physics and Interdisciplinary Research, Soochow University, Suzhou 215006, China}

\date{\today}

\begin{abstract}
We study a robust model of self-propelled rods interacting via volume exclusion and
show that its collective dynamics encompasses both that of the corresponding Vicsek-style model (where local alignment is the sole interaction),
and motility-induced phase separation (which occurs when repulsion is dominant).
These results unify these heretofore largely disconnected bodies of knowledge on dry active matter and clarify the nature of the various phases involved.
\end{abstract}

\maketitle

Active matter systems, where particles use energy to move persistently, are nowadays routinely divided into dry and wet ones.
Dry situations are those where the fluid displaced by the particles' motion can be neglected, which is often the case
for particles crawling, walking, sliding on a substrate such as some bacteria
\cite{MYXO,Zhang2010}, biofilaments in motility assays \cite{SUMINO},
granular particles moving on a vertically-shaken horizontal plane
\cite{DESEIGNE1,DESEIGNE2,Kudrolli2008,SRIRAM1,SRIRAM2}, etc.

Our theoretical understanding of dry active matter is currently somewhat piecemeal.
It is quite satisfactory
when local alignment is the dominating interaction, such as in Vicsek-style models,
where constant-speed point particles locally align their velocities.
There a general scenario has emerged
in terms of a phase separation between a disordered gas and an orientationally ordered liquid endowed with generic long-range correlations and anomalously strong number fluctuations \cite{AIM,SCT}.
When steric repulsion is dominating and no significant alignment exists, such as for active Brownian particles (ABPs),
the situation is also satisfactory. The most striking phenomenon exhibited by these systems is motility-induced phase separation (MIPS),
which arises in spite of the absence of explicit attractive interactions
\cite{MIPS1a,MIPS1b,MIPS1c,MIPS1d}.
Our understanding of MIPS is now rather complete,
and the first steps towards a thermodynamics of this class of active matter have been made
\cite{MIPS2a,MIPS2b,MIPS2c,MIPS2d,MIPS2e}.
These two sets of results can be thought of as limit cases. Even if the Vicsek- and the MIPS-worlds have some experimental relevance
\cite{DAIKI,SRIRAM2,palacci2013,cottin2012},
the more generic situation is when both alignment and repulsion play a role, as with self-propelled elongated objects interacting by volume exclusion: in the typically overdamped framework of most dry active matter systems, the inelastic collision of two rods leads to (nematic) alignment.
Our current knowledge of such systems is far less satisfactory, and moreover their reported collective properties seem largely disconnected from those of Vicsek-like and MIPS systems.

This is true in particular in the case of overdamped self-propelled {\it polar} rods (without velocity reversals),
one of the earliest dry active matter models, investigated in the general context of crawling and sliding bacteria
\cite{Peruani2006}.
Works on such self-propelled rods (SPR, from now on) have revealed
a relatively rich set of collective phenomena
\cite{rodWensink1,rodWensink2,rodPeruani,rodBetterton,rodHagan,rodYang,rodAbkenar},
but they do not include the fluctuating active nematic state nor the  segregated
nematic bands characteristic of the coexistence phase of the Vicsek-style rods model
\cite{VICSEK-RODS-MICRO,VICSEK-RODS-HYDRO}.
Also, MIPS typically vanishes from microscopic models such as ABPs as soon as one considers even slightly anisotropic particles
\footnote{Behavior reminiscent of MIPS has been reported for elongated objects \cite{LETICIA,rodYang,rodHagan,rodPeruani}, but these works remain disconnected from the bulk of MIPS studies.}

Moreover, in our opinion, the asymptotic status of the two dominating non-trivial regimes observed in SPR,
{\it i.e.} moving polar clusters and nematic laning (see Fig.~\ref{fig3}b,d), is not well established (see below).
Finally, only a partial exploration of an a priori rather large parameter space is typically presented,
so that a global picture of the phase diagram of SPR systems remains unavailable.

In this Letter, we study a robust and efficient SPR model and we show that its collective dynamics
encompasses both the Vicsek- and the MIPS-worlds,
once one varies the anisotropy of friction and the softness of the repulsive potential at play between rods.
Our results thus unify heretofore largely disconnected bodies of knowledge on dry active matter.
While the detailed and complete exploration of the phase diagram of our model will be presented in \cite{PRE},
here we focus on some of the most salient results obtained:
(i) partial phase diagrams showing how Vicsek- and MIPS- phenomena are connected to more standard SPR phases;
(ii) a detailed study revealing that the polar clusters regime is a {\it bona fide} micro-phase separated state;
(iii) evidence that the nematic laning regime is actually coarsening at long times and ultimately unstable in large enough systems;
(iv) existence of several types of dense phases within the MIPS regime.

\begin{figure}[t!]
\includegraphics[width=\columnwidth,clip=on]{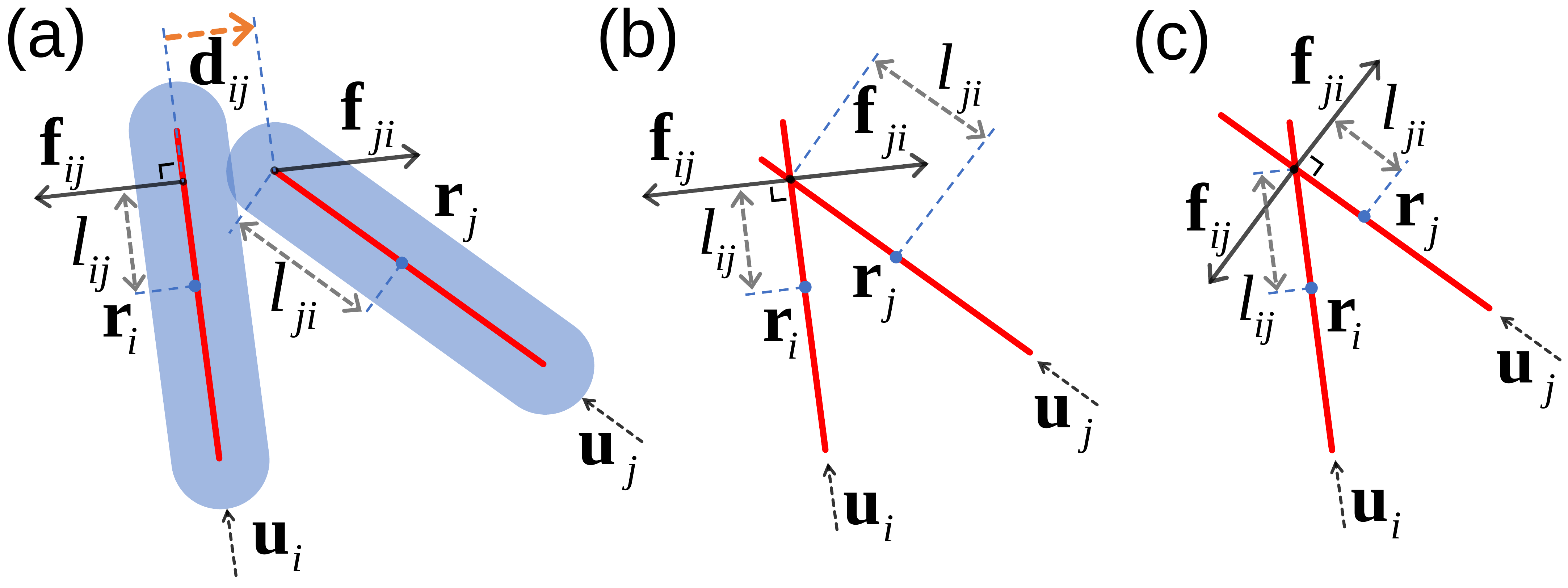}
\caption{Schematics of the interaction of rods $i$ and $j$ of polarity ${\bf u}_i$ and ${\bf u}_j$.
(a) Interaction without overlap: rod $j$'s tip collides with rod $i$'s body and they feel opposite forces ${\bf f}_{ij}$ and ${\bf f}_{ji}$  given
by the orange vector ${\bf d}_{ij}$, perpendicular to $i$,
that links the point on the red backbone of $i$ closest to the end of the red backbone of $j$.
(b,c) Treatment of overlaps: Once $j$'s backbone reaches $i$'s backbone
($|{\bf d}_{ij}|=0$) and then crosses it, ${\bf f}_{ij}$ and ${\bf f}_{ji}$ are kept constant
as long as the segment $j$'s backbone crossing that of $i$ remains the shortest
of the 4 segments defined by the crossing backbones (b).
If this shortest segment changes to belong to $i$'s backbone,
then the forces rotate to become perpendicular to ${\bf u}_j$ (c).
At all times, the shortest segment determines the orientation of the forces.}
\label{fig1}
\end{figure}

\begin{figure*}[t!]
\includegraphics[width=0.9\textwidth,clip=on]{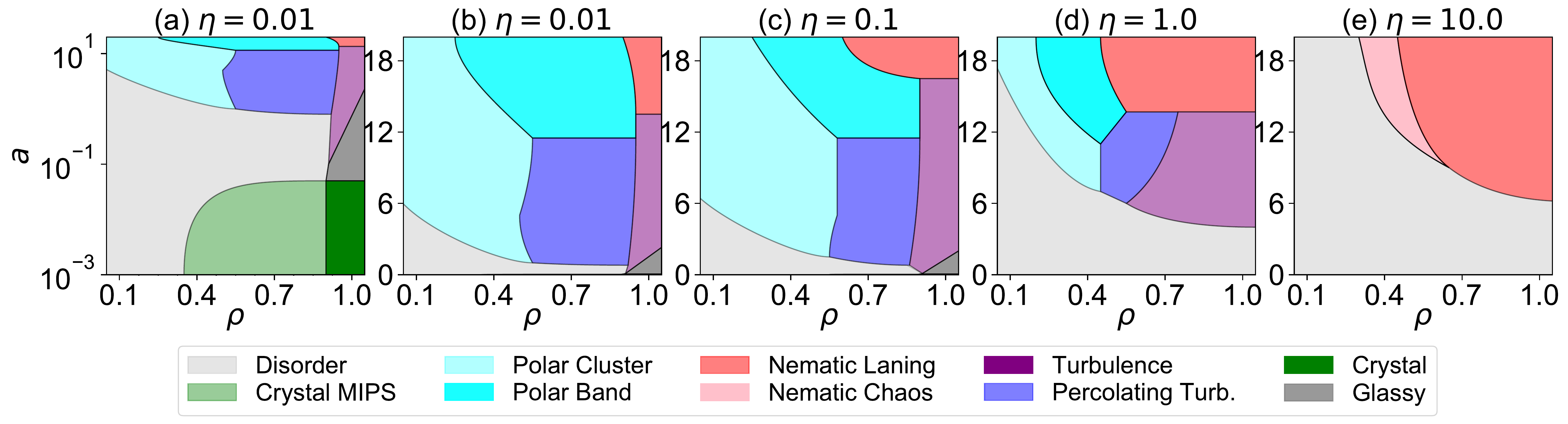}
\caption{Phase diagram at `standard parameters' (see text) and hard repulsion ($S=10^{-2}$)
obtained from steady-state simulations of 4000 rods in square domains
with periodic boundary conditions,
following an initial condition with nematic order and random positions (phases defined in Table~\ref{table}, illustrated in Fig.~\ref{fig3}).
(a-e): packing fraction/aspect ratio $(\rho,a)$ planes at various noise values $\eta$ as indicated on top of panels.
Note that panels (a) and (b) are both for $\eta=0.01$, with (a) using a log scale to reveal the crystal MIPS region at small $a$.}
\label{fig2}
\end{figure*}

{\it A robust and efficient SPR model.} Even in the restricted setting of dry, overdamped two-dimensional systems, many factors define a SPR model: aspect ratio of rods, softness of potential,
friction/motility tensor, noise strengths, packing fraction. Details in the {\it shape} also matter \cite{shape}.
Past works have used ellipsoids, needles, rectangles
\cite{Peruani2006,needle,ellipse}.
Particularly popular are rods made of rigidly connected, overlapping balls\cite{rodPeruani,rodWensink1,rodWensink2,rodAbkenar,rodYang}.
This allows to calculate the rod-rod interaction as a sum of pairwise forces between balls, a simple solution,
but a rather inefficient one in the limit of high aspect ratio.
Past works also often use rather hard repulsion potentials\cite{rodBetterton,rodYang},
another numerically inefficient limit, and zero or weak noise\cite{rodWensink1,rodWensink2},
which accentuates the influence of sharp angles (as with rectangles) and non-convex shapes (as with chains of beads).

\begin{table}[t!]
\caption{\label{table}
Collective states appearing in the phase diagrams.
First row: quantifiers used to define them:
$\phi_6=\langle|\langle\exp(6 {\rm i} \theta_{j,j+1})\rangle_{j\sim i}|\rangle_i$, where $\theta_{j,j+1}$ is the bond angle joining two consecutive
Voronoi neighbors of $i$ and the inner summation is performed over Voronoi neighbors. Local polar and nematic order parameters
$p$ and $q$ are similarly defined using Voronoi neighbors. Their global counterparts are $P$ and $Q$.
PS is an indicator of density segregation (usually detected using the method described in \cite{SI}),
either in a single, large dense structure (macro) or in finite-size dense objects (micro).
Symbol $\circ$ (resp. $\times$): the presence (resp. absence) of these quantifiers.
}
\begin{ruledtabular}
\begin{tabular}{llcccccc}
& & $\phi_6$ & $p$ & $q$ & $P$ & $Q$ & PS \\
\multicolumn{2}{l}{disorder} & $\times$ & $\times$ & $\times$ & $\times$ & $\times$ & $\times$ \\
\multicolumn{2}{l}{crystal}  & $\circ$ & $\times$ & $\times$ & $\times$ & $\times$ & $\times$ \\
\multicolumn{2}{l}{turbulence}  & $\times$ & $\circ$ & $\circ$ & $\times$ & $\times$ & $\times$ \\
\multicolumn{2}{l}{nematic}  & $\times$ & $\times$ & $\circ$ & $\times$ & $\circ$ & $\times$  \\
\multirow{3}{*}{nematic$\Bigg\{$} &\hspace{-0.2cm} laning & $\times$ & $\circ$ & $\circ$ & $\times$ & $\circ$ & $\times$ \\
                                      & \hspace{-0.2cm} chaos & $\times$ & $\times$ & $\circ$ & $\times$ & $\times$ & $\times$ \\
                                      &\hspace{-0.2cm} band & $\times$ & $\times$ & $\circ$ & $\times$ & $\circ$ & macro \\
\multirow{3}{*}{MIPS$\Bigg\{$} & \hspace{-0.4cm}glass/liquid & $\times$ & $\times$ & $\times$ & $\times$ & $\times$ & macro \\
                                   & \hspace{-0.5cm} turbulent & $\times$ & $\circ$ & $\circ$ & $\times$ & $\times$ & macro \\
                                   & \hspace{-0.4cm}crystal & $\circ$ & $\times$ & $\times$ & $\times$ & $\times$ & macro \\
\multicolumn{2}{l}{polar clusters} & {$\times$} & {$\circ$} & {$\circ$} & {$\times$} & {$\times$} & {micro} \\
\multicolumn{2}{l}{polar band} & $\times$ & $\circ$ & $\circ$ & $\circ$ & $\circ$ & macro \\
\end{tabular}
\end{ruledtabular}
\end{table}

Here we consider spherocylinders of width $d_0$ and total length $(a+1)d_0$,
a geometry that allows to reduce their interaction to that between 2 points (Fig.~\ref{fig1}a)
\cite{rodfluid1,rodfluid2}.
We choose harmonic repulsion and we generally stay away from the zero-noise limit, all features which, combined with the convex and smooth shape
of our rods, guarantee robustness and efficiency. The equations governing the position ${\bf r}_i$ of the center of mass of rod $i$
and its orientation $\theta_i$ read:
\begin{align}
\partial_t {\bf r}_i &= \mu_\| s_0 {\bf u}_i + {\bf \Pi}_i \sum_{j\neq i} {\bf f}( {\bf d}_{ij} ) + \boldsymbol{\eta}_i \\
\partial_t \theta_i &= \mu_{\theta} \sum_{j\neq i}  [ l_{ij}{\bf u}_i \times {\bf f}( {\bf d}_{ij})] \cdot {\bf z}
+ \eta_i^\theta
\end{align}
with {\bf z} the unit vector perpendicular to the plane of motion and the force ${\bf f}$ given by
\begin{equation}
{\bf f}( {\bf d}_{ij}) = k H(d_0-|{\bf d}_{ij}|) \frac{{\bf d}_{ij}}{|{\bf d}_{ij}|}
\end{equation}
where
${\bf u}_i$ is the unit vector along $\theta_i$,
${\bf d}_{ij}$ and $l_{ij}$ are defined in Fig.~\ref{fig1}a,
$H$ is the Heaviside step function, $ {\bf \Pi}_i = \mu_\| {\bf u}_i  {\bf u}_i + \mu_\perp ({\bf I} - {\bf u}_i  {\bf u}_i ) $ is the mobility tensor,
$\boldsymbol{\eta}_i = \sqrt{2\eta\mu_\|} \xi^\|_i {\bf u}_i + \sqrt{2\eta\mu_\perp} \xi^\perp_i  ({\bf u}_i \times {\bf z})$ is an anisotropic noise,
$\eta_i^\theta = \sqrt{2\eta\mu_\theta} \xi^\theta_i$,
and $\xi^\alpha_i$ are Gaussian white noises with unit variance.
From the self-propulsion force intensity $s_0$ and the spring constant $k$, we define a softness coefficient $S=s_0/k$.
Except for very low softness, our rods can overlap and cross each other, a feature that is actually realistic when dealing with quasi-2D situations
such as some bacteria swarming films and motility assays. Our treatment of overlaps is detailed in Fig.~\ref{fig1}b,c.

\begin{figure}[b!]
\centerline{\includegraphics[width=\columnwidth,clip=on]{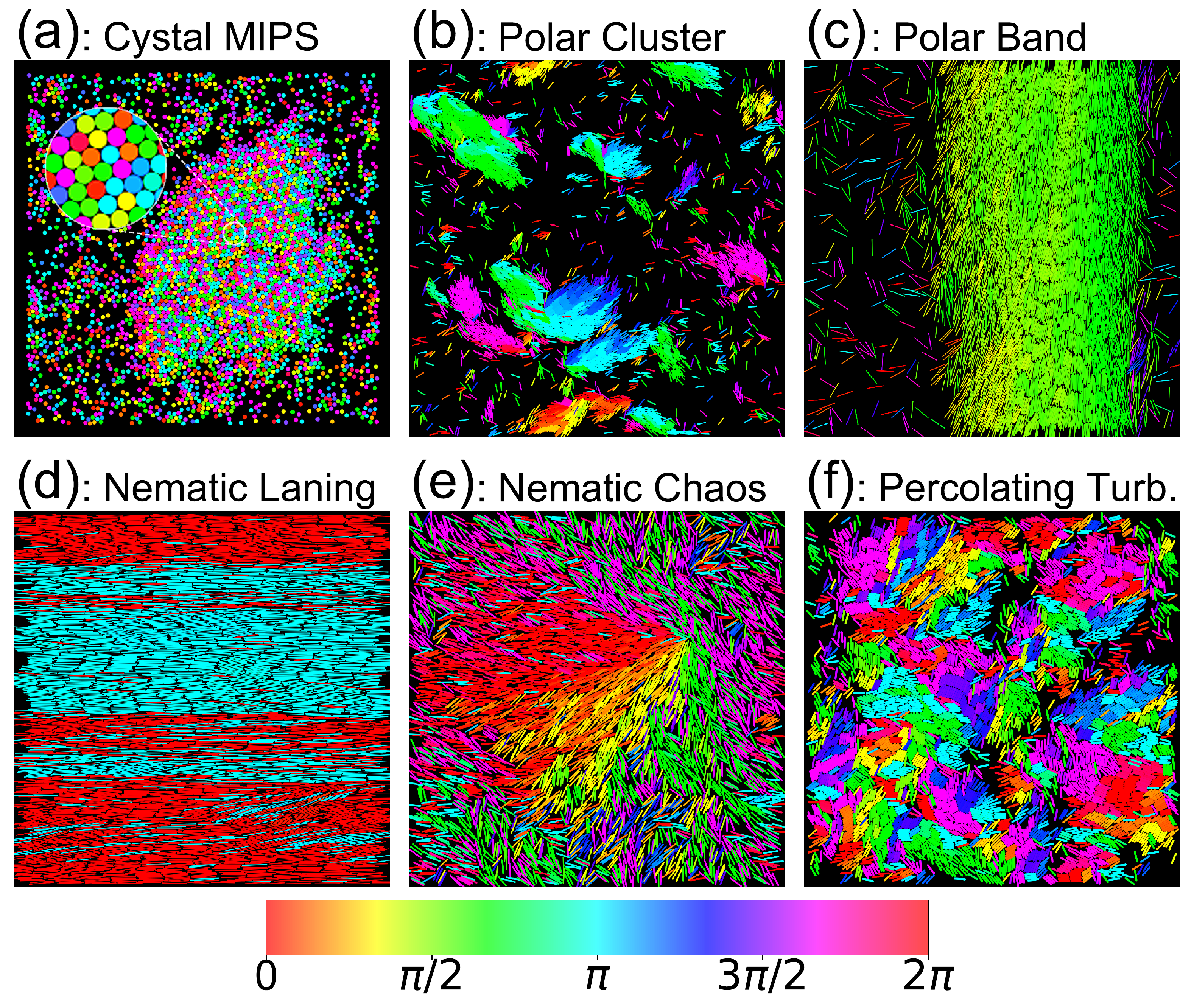}}
\caption{Snapshots of main phases reported in Fig.~\ref{fig2} ($N=4000$, $S=10^{-2}$).
Rods are colored by their polarity angle.
(a) crystal MIPS in ABP limit ($a=0$, $\rho=0.5$ and $\eta=0.01$); a large, immobile dense cluster has formed.
Inset: zoom showing local crystalline order.
(b) polar clusters moving coherently in various directions amidst a gas ($a=10$,  $\rho=0.2$  and $\eta=0.1$),
(c) polar band (here the rods in the band move "upward") ($a=20$, $\rho=0.4$  and $\eta=1$),
(d) nematic laning (here with 2 main polar lanes inside which rods move in opposite directions) ($a=17.5$, $\rho=1$  and $\eta=1$),
(e) nematic chaos (in this very dynamic state, $+\frac{1}{2}$ defects such as the one in the upper right corner move and reorganize constantly
the system) ($a=10$, $\rho=0.6$  and $\eta=10$),
(f) percolating turbulence (locally aligned rods move constantly, forming a single, almost space-covering, globally-disordered cluster) ($a=5$, $\rho=0.7$  and $\eta=0.1$).
}
\label{fig3}
\end{figure}

\begin{figure}[b!]
\includegraphics[width=\columnwidth,clip=on]{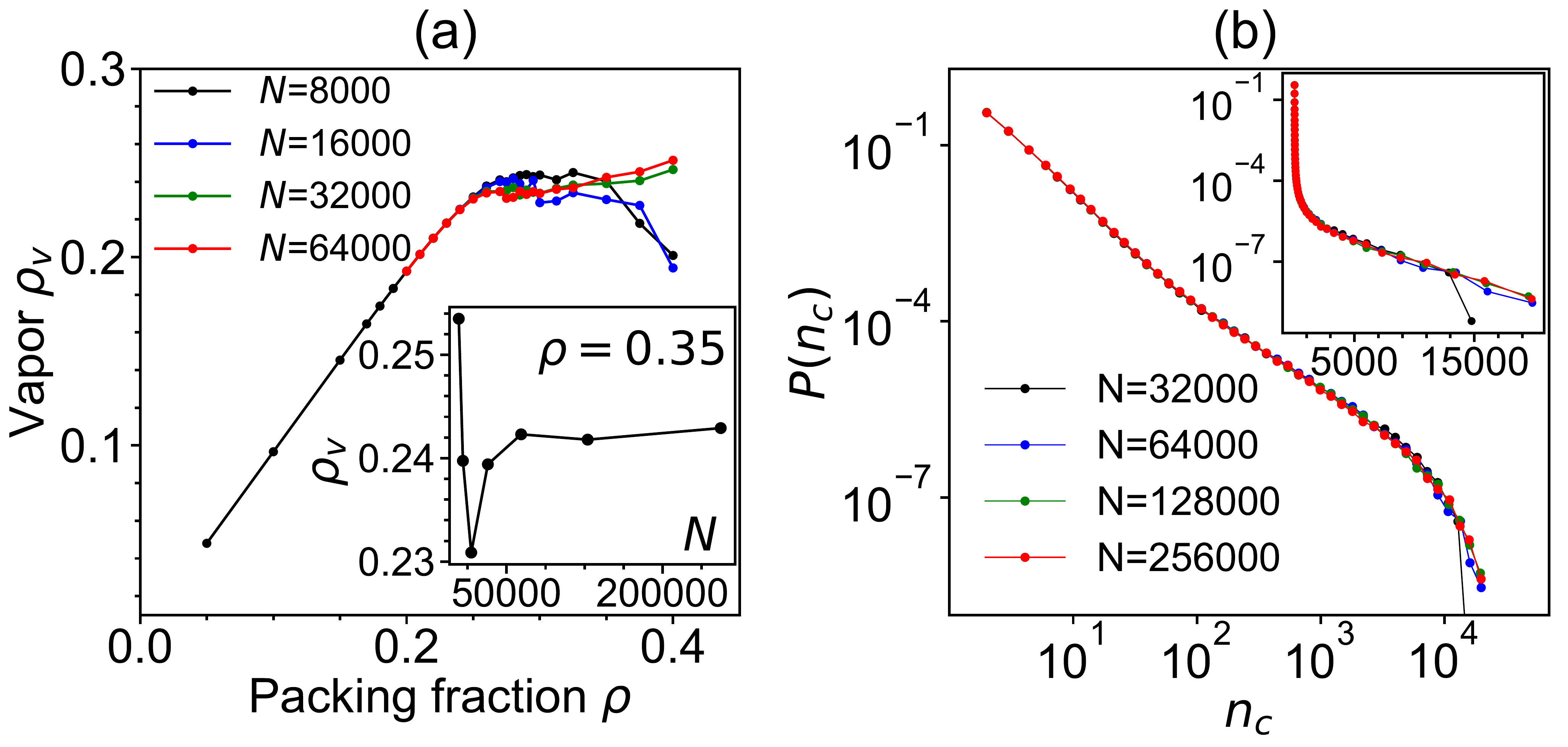}
\caption{The polar clusters phase is microphase separated.
(a) Vapor density vs global packing fraction for different system sizes across the gas/polar clusters phase transition.
(b) Cluster size distribution at fixed parameters for different system sizes.
Inset: zoom on the tail in linlog scales.
In both panels $a=10$, $\eta=1$, $S=10^{-1}$. In (b) $\rho=0.35$.
}
\label{fig4}
\end{figure}

{\it Phase diagram at `standard' parameters.}
Most existing work on SPR has been performed
using ratios of motility coefficients dictated by those of a rod
in an equilibrium bath, with some variations \cite{MOTILITY}.
Here our `standard' choice is $\mu_\|=\mu_\perp$, $\mu_\theta/\mu_\perp=12/[1+(1+a)^2]$.
With these ratios fixed, choosing without loss of generality $d_0=s_0=\mu_\|=1$, the model has 4 free parameters:
softness $S$, aspect ratio $a$, packing fraction $\rho$, and P\'eclet number or noise strength $\eta$.
Previous works are in the hard repulsion limit ($S$ small) and only vary a subset of the 3 remaining parameters.
Here this complete 3D phase diagram, obtained at $S=10^{-2}$ for $N=4000$ rods, is presented in Fig.~\ref{fig2}.
Its main features are in agreement with previous partial observations (see, e.g., \cite{rodBetterton,rodYang,rodAbkenar}).
Its detailed structure, and in particular
the full-coverage phases present at large $\rho$ values, will be discussed in \cite{PRE}.
We focus on the general interplay between the MIPS, polar clusters, polar band, percolating turbulence, and nematic chaos and laning phases,
snapshots of which can be found in Fig.~\ref{fig3}, together with a brief description of their dynamics (see also Movies~1-6 in \cite{SI}).
These 6 phases are naturally grouped into 3 sets:
(i) polar clusters, single polar band, and percolating turbulence (blueish colors) have local polar order;
(ii) nematic laning and chaos have local nematic order (redish colors);
(iii) in the green MIPS region, which is confined to low noise
and very small aspect ratio
(Fig.~\ref{fig2}a), the dense macroscopic cluster has local crystalline order.
The situation described in the introduction is confirmed:
even slightly elongated particles cannot show MIPS, which is only connected to isotropic disorder.
The regimes characteristic of the Vicsek-style rods model (homogeneous nematic state and segregated nematic bands) are not present.

Whereas determining the asymptotic (infinite-size) phase diagram is a difficult task beyond the scope of this paper,
it is useful to point to simplifications occurring in this limit.
We observed that the opposite-going polar bands forming the laning regime slowly coarsen in time;
in wide-enough systems, enough compression occurs, resulting in sparse areas where fluctuations
eventually generate small, transversely-moving clusters that destroy the pattern, leaving nematic chaos (Movie~7 in \cite{SI})
\footnote{Note that this has been reported briefly also in \cite{rodHagan}}.
Increasing system size, we find that the polar band region gradually disappears,
leaving a larger region of polar clusters
\footnote{In other words, polar bands are just polar clusters large-enough to span one dimension of the system,
as suggested in \cite{rodPeruani}, and not a separate phase, as suggested in \cite{rodAbkenar,rodBetterton}}.
We performed a detailed analysis of this dominating polar clusters phase.
One difficulty is to find a proper definition of clusters.
Usually, this is done via a criterion on some measure of the local density $\rho_i$ around rod $i$,
since one can expect a bimodal distribution of $\rho_i$ in the presence of clusters.
Here, we use instead a criterion on the local polar order $p_i$ defined via a Voronoi tesselation adapted to elongated objects,
a procedure we detail in \cite{SI}.
Using such a criterion, we find that when increasing the overall packing fraction,
the vapor density stops growing and remains constant as the polar clusters appear (Fig.~\ref{fig4}a).
One is thus in presence of a phase separation scenario. The dense phase is made of the polar clusters.
They are well-defined {\it microphases}: their size distribution may appear powerlaw-like at moderate sizes but converges
to a well-bounded form for large-enough systems; there exists a well-defined maximal cluster size (Fig.~\ref{fig4}b;
a very large system is shown in \cite{SI}, Fig.~S1c and Movie 8).

We have thus clarified the status of the two emblematic phases of SPR in the infinite-size limit: nematic laning seems intrinsically unstable,
and polar clusters are the finite-size microphases of a bona fide density-separated phase.
Extrapolating these results, we conclude that the polar clusters and the nematic chaos phases
are probably the only regimes left asymptotically (in addition to MIPS and the disordered gas).

\begin{figure}[t!]
\includegraphics[width=\columnwidth,clip=on]{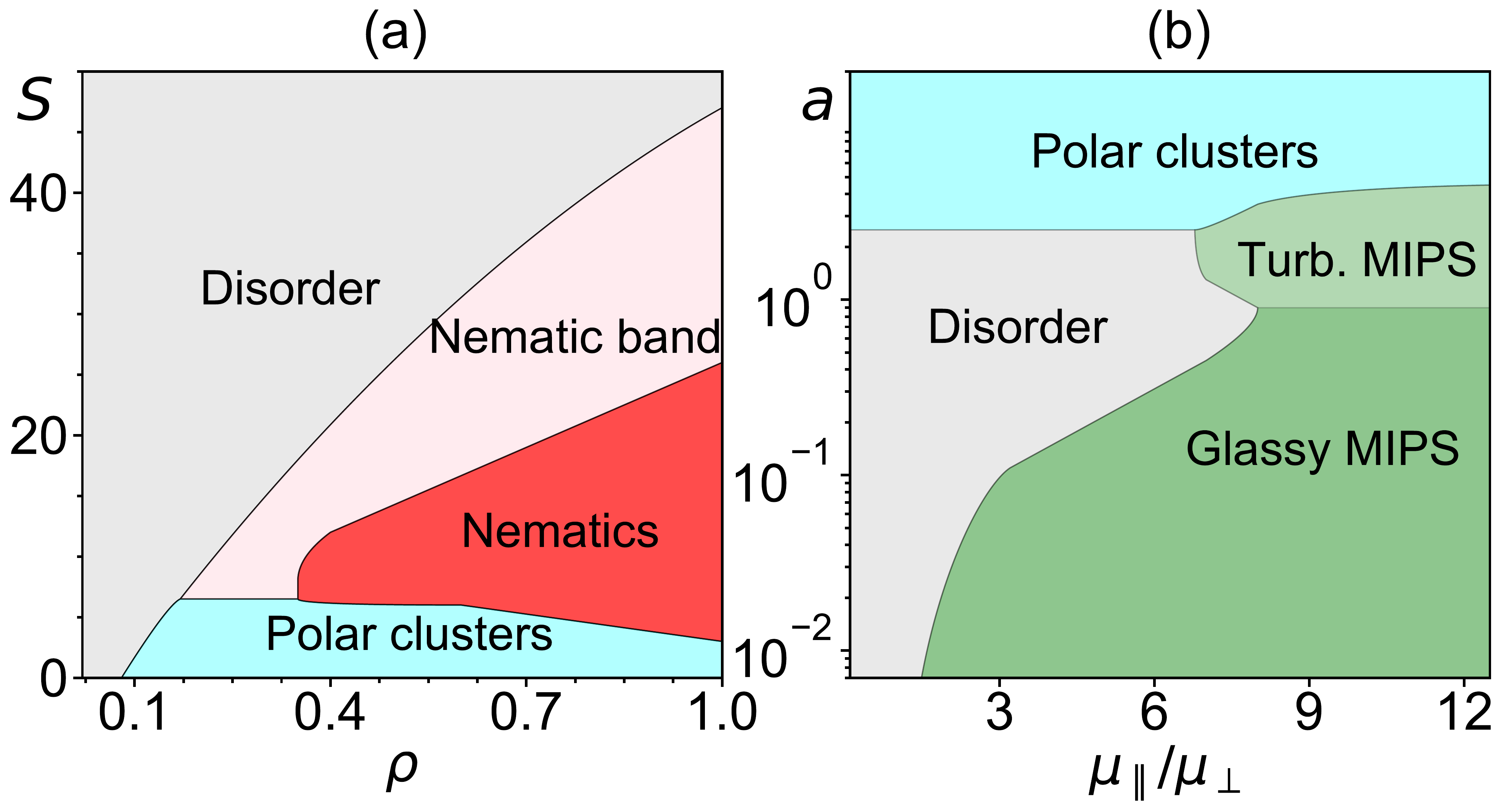}
\caption{
(a) $(\rho,S)$ phase diagram at $\eta=0.1$, $a=11.5$, and `standard' motility coefficients.
(b) $(\mu_\|/\mu_\perp, a)$ phase diagram at $\eta=0.01$, $\rho=0.3$, $S=0.01$. $N=8000$ in both panels.
}
\label{fig5}
\end{figure}

{\it Connecting the polar clusters phase to Vicsek and MIPS phenomenology.}
As noted above, the `standard' 3D phase diagram does not include the dense nematic bands
characterizing the coexistence phase of the Vicsek-style rods model nor its homogeneous nematic phase.
However, they can both be observed when increasing $S$, the softness of the repulsive potential,
as shown by a phase diagram in the $(\rho,S)$ plane at fixed noise and aspect ratio (Fig.~\ref{fig5}a, Movie 9 in \cite{SI}).
For $S$ large enough, overlaps between rods are made easier, and alignment less systematic,
allowing the emergence of a Vicsek-style homogeneous nematic phase with long-range correlations and
giant number fluctuations (not shown)
\footnote{This is similar to the nematic phase observed experimentally in \cite{DAIKI}
where elongated E. coli cells weakly align upon collision but can also frequently overlap.}.

\begin{figure}[t!]
\includegraphics[width=\columnwidth,clip=on]{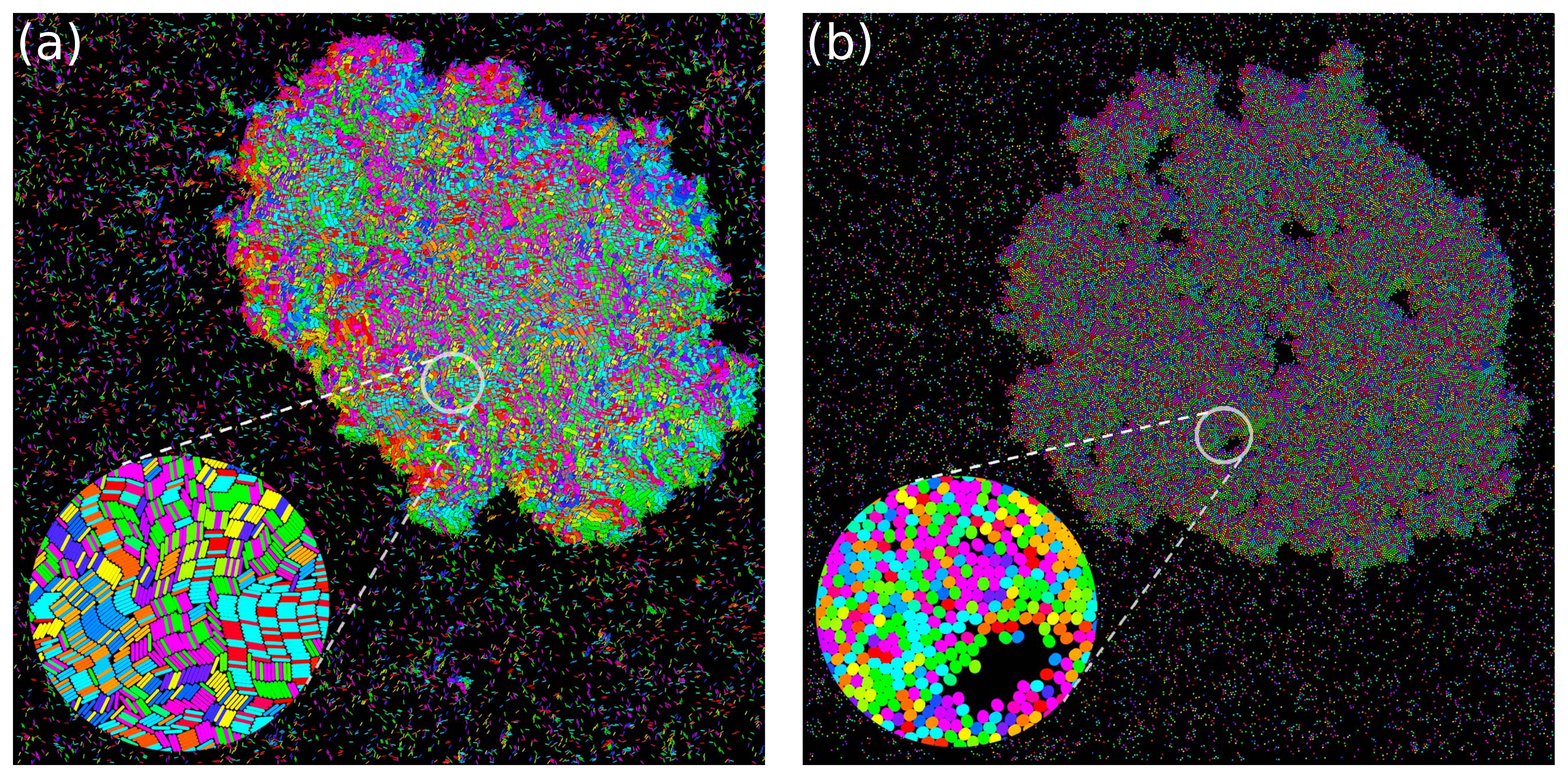}
\caption{Typical snapshots in the two MIPS regimes reported in Fig.~\ref{fig5}b.
Colors as in Fig.~\ref{fig3}, $\mu_\|/\mu_\perp=10$, $N=64000$.
Insets are zooms allowing to see local structure.
(a) turbulent MIPS $a=3.5$.
(b) glassy MIPS with bubbles $a=0.2$.
}
\label{fig6}
\end{figure}

As also noted above, the MIPS region disappears even for weakly-elongated particles,
leaving only the gas phase (Fig.~\ref{fig2}a). This observation, made at standard parameters, persists
when $S$ is increased. To connect MIPS to the dominating polar clusters phase exhibited
by long rods, one needs to break away from the `equilibrium'-dictated ratios between $\mu_\|$, $\mu_\perp$, and $\mu_\theta$ used so far.
(Note that this is something quite legitimate in the dry active matter context, with elongated objects in contact with a substrate.)
Increasing the friction perpendicular to the rod's axis and the rotational friction (i.e. decreasing $\mu_\perp$ and $\mu_\theta$
keeping $\mu_\theta/\mu_\perp$ as before), a direct transition from MIPS to polar clusters can occur. This transition can be governed
by $a$, the aspect ratio of rods, as shown in the $(\mu_\|/\mu_\perp, a)$ phase diagram of Fig.~\ref{fig5}b. Note that near this transition
MIPS takes place for rather long rods, something never reported before for smooth objects
\footnote{In \cite{rodPeruani}, macroscopic clusters made of long rods were also reported.
However, our model does {\it not} show MIPS at the same parameters.
We believe that the non-concave shape of the rods made or overlapping disks used in \cite{rodPeruani} are at the origin
of the phase separation reported there.}.
The small values of $\mu_\perp$ and $\mu_\theta$ are such that a rod colliding into another is often stopped on its course
instead of rotating and gliding along the other, favoring the formation of blocked clusters, without resorting to a wedge geometry as in \cite{KUMAR}.
The resulting macroscopic clusters are, however, rather dynamic (`turbulent' MIPS, Fig.~\ref{fig6}a and Movie 10 in \cite{SI}).
Rods are also strongly aligned locally inside them.
All this is in contrast to the `glassy bubbly MIPS', disordered dense phase present for shorter rods (Fig.~\ref{fig6}b and Movie 11 in \cite{SI}).
If one now remarks that, in the MIPS region of Fig.~\ref{fig2}, the dense phase has local crystalline order,
then one must in principle distinguish at least three MIPS regions. Whether the nature of the dense phase in MIPS makes
any difference at large-scales is the subject of ongoing work.

To summarize, self-propelled rods interacting via volume exclusion can display
both the phenomenology of Vicsek-style models with polar particles aligning nematically
and that of motility-induced phase separation, which is thus not limited to isotropic particles anymore.
To achieve this unification of dry active matter phenomena, we have shown that one needs to vary 5
parameters: in addition to those usually considered (rod aspect ratio, packing fraction, and noise/P\'eclet number),
the softness of the repulsive potential (offering the possibility of overlaps) and the anisotropy of motility/friction play a qualitative role.

\acknowledgments
We thank Eric Bertin, Francesco Ginelli, Beno\^{\i}t Mahault, and Cesare Nardini for a critical reading of this manuscript.
This work is partially supported by ANR project Bactterns, FRM project Neisseria,
and the National Natural Science Foundation of China (grants \#11635002 to X.-q.S. and H.C., \#11474210 and \#11674236 to X.-q.S.).

\end{document}